\documentclass[oldversion, printer]{aa}
\usepackage[T1]{fontenc}
\usepackage{psfig}
\usepackage{subfigure}
\usepackage{graphicx}
\usepackage{latexsym}
\usepackage{mathrsfs}
\usepackage{amsmath}
\usepackage{amssymb}
\usepackage{amsfonts}
\usepackage{multirow}
\usepackage{ulem}
\usepackage{natbib}
\usepackage[usenames,dvipsnames]{color}

\advance \voffset by 1.00cm\relax

\newcommand{\der}[2]{\ensuremath{\frac{{\rm d} #1}{{\rm d} #2}}}

\newcommand{\pder}[2]{\ensuremath{\frac{\partial #1}{\partial #2}}}
\newcommand{\kpder}[3]{\ensuremath{\frac{\partial #1}{#2 \partial #3}}}
\newcommand{\npder}[3]{\ensuremath{\frac{{\partial}^{#3} #1}{\partial #2^{#3}}}}

\newcommand{\pderln}[2]{\ensuremath{\frac{\partial\,\rm ln\,#1}{\partial\,\rm ln\,#2}}}

\def\spose#1{\hbox to 0pt{#1\hss}}
\def\lta{\mathrel{\spose{\lower 3pt\hbox{$\mathchar"218$}}
         \raise 2.0pt\hbox{$\mathchar"13C$}}}
\def\gta{\mathrel{\spose{\lower 3pt\hbox{$\mathchar"218$}}
         \raise 2.0pt\hbox{$\mathchar"13E$}}}
\newcommand{\be}{\begin{equation}}
\newcommand{\ee}{\end{equation}}
\newcommand{\bea}{\begin{eqnarray}}
\newcommand{\eea}{\end{eqnarray}}



\begin{document}
%=================================================================

\title{Stability of radiation-pressure dominated disks.\\ I. The dispersion relation for a delayed heating $\alpha$--viscosity prescription}
%=================================================================
%   \subtitle{}
%=================================================================
    \author{
           Adam Ciesielski \inst{1,2}
           \and
           Maciej Wielgus \inst{3}
           \and
            W{\l}odek Klu{\'z}niak\inst{4}
           \and
              Aleksander S\k{a}dowski \inst{4}
           \and
               Marek Abramowicz\inst{4,5}
          \and\\
              Jean-Pierre Lasota\inst{1,6}
           \and
              Paola Rebusco\inst{7}
               }
% %=================================================================
   \institute{
             Astronomical Observatory of the Jagiellonian University, ul. Orla 171,
             PL-30-244 Krak{\'o}w,
             Poland\\
             \email{adam.ciesielski@uj.edu.pl}
             \and
             M. Smoluchowski Institute of Physics, Jagiellonian
             University, ul. Reymonta 4,
             PL-30-059 Krak{\'o}w,
             Poland\
             \and
Institute of Micromechanics and Photonics, Warsaw University of Technology, ul.
             \'Sw. Andrzeja Boboli 8, PL-02-525 Warszawa, Poland\\
             \email{maciek.wielgus@gmail.com}
             \and
             Nicolaus Copernicus Astronomical Center, Polish Academy
             of Sciences,
             ul. Bartycka 18, PL-00-716 Warszawa, Poland \\
             \email{wlodek@camk.edu.pl}, ~\email{as@camk.edu.pl}
             \and
             Department of Physics, G\"oteborg University,
             SE-412-96 G\"oteborg, Sweden    \\
             \email{marek.abramowicz@physics.gu.se}
             \and
             Institut d'Astrophysique de Paris, UMR 7095 CNRS, UPMC Univ Paris 06, 98bis~Boulevard Arago, 75014 Paris, France\\
             \email{lasota@iap.fr}
             \and
             Experimental Study Group, MIT, Cambridge, MA 02139, USA\\
             \email{pao@space.mit.edu}
            }
%=================================================================
%=================================================================
\date{Received ????; accepted ???? }
%=================================================================
%=================================================================
\abstract{We derive and investigate the dispersion relation for accretion disks with retarded or advanced heating. We follow the $\alpha$-prescription but allow for a time offset $\tau$ between heating and pressure perturbations, as well as
for a diminished response of heating to pressure variations. We study in detail solutions of the dispersion relation for disks with radiation-pressure fraction $1-\beta$. 
%We show that negative real roots exist for $\Omega\tau\ge 0$ if $\xi$ is sufficiently small --- for $\beta=0$ it has to satisfy $\xi<1/2$. 
For $\tau< 0$ (delayed heating) the number and sign of real solutions  for the growth rate depend on the values of the time lag and the ratio of heating response to pressure perturbations, $\xi$. If the delay is larger than a critical value (e.g., if $\Omega\tau<-125$ for $\alpha=0.1$, $\beta=0$ and $\xi=1$) 
two real solutions exist, which are both negative. These results imply that retarded heating may stabilize radiation-pressure dominated accretion disks.}
%=================================================================
\authorrunning{A. Ciesielski et al.}
\titlerunning{Dispersion relation for retarded and advanced heating}
%=================================================================
\keywords{black holes physics --- accretion disks --- stability ---
time-delay} \maketitle
%=================================================================

\section{Introduction}
%X-ray binaries and AGN exhibit different types of variability %taking place at different timescales. Some of them may result %from intrinsic instabilities of accretion disks. Other may %result, e.g., from the orbital motion. Instabilities caused by %the viscous and non-adiabatic processes in disks may be %classified into three types \citep{bb}: thermal, secular %instabilities and overstabilities of waves and oscillations. %The stability of accretion disks against the first two has %been studied since the middle of the 1970's %\citep[e.g.,][]{shibazakihoshi-75, %shakurasunyaev-76,pringle-76}. It was proven that the standard %$\alpha$-prescription leads to unstable radiation-pressure %dominated disks.
%However, despite the fact that most of observed galactic BHs %happen to accrete matter at high accretion rates, they do not %exhibit oscillations on timescales which could be associated %with the thermal instability. This inconsistency has remained %unsolved.

The \citet{shakura-73} $\alpha$\,-viscosity accretion disk model has been extremely successful in describing various astronomical objects and systems. The only exception is its application to systems accreting at high rates. At rates where pressure is dominated by that of radiation and opacity by electron scattering, the $\alpha$ disk is thermally and viscously (secularly) unstable \citep{LE74,shakurasunyaev-76,shibazakihoshi-75}. In the case of accretion onto black holes this regime corresponds to luminosities in excess of $\gta 0.01\, L_{\rm Edd}$, where $L_{\rm Edd}=GMm_p\,c/\sigma_T$, $m_p$ is the proton mass and $\sigma_T$ the Thompson cross-section. However, black-hole X-ray sources cross this limit upwards to maximum luminosity and downwards to minimum luminosity showing no dramatic symptoms at all \citep[but they enter the so-called hard/low state, see e.g., ][]{mcr06}, and certainly not the behavior anticipated by models \citep[e.g.,][]{lp91,TL}. Observations suggest that disks in black-hole transient systems are stable up to at least $\sim 0.5\,L_{\rm Edd}$ \citep{dwg}.

Models of radiation-pressure dominated disks are unstable only when the viscous stress is proportional to the total pressure ($\alpha$ being the proportionality constant).
Models with ad hoc viscosity prescriptions have been studied
but their relation to reality remains unknown. Simulations of the magneto-rotational instability in radiation-pressure dominated disks \citep[see e.g.,][]{turner04} showed that stress was approximately proportional to the total pressure, but they exhibited no sign of instability.

Recently, \cite{hirose-09-1} showed that although the linear correlation of vertically integrated stress and pressure is roughly satisfied in shearing-box MHD simulations of radiation-pressure dominated disks, these quantities are shifted in time---pressure responds to stress variations after $\sim10-20$ dynamical times.
Using these simulation results as a guideline, we perform an analytical, perturbative study of the stability of disks with a modified viscosity prescription that allows for such a time lag between stress and pressure. 

The theory of delayed oscillators has already been extensively developed \citep[e.g.,][]{minorsky-44,cookegrossman-82,bellmancooke-63}. It predicts that an oscillator may be easily stabilized (or destabilized) if only the system parameters are chosen properly. We show that this is the case also for accretion disks. Similar conclusions were recently obtained in another analytic study parallel to this work by \cite{linetal-11}.

This work is devoted to the mathematical part of this project. In Sect.~\ref{s.prescription} we discuss our choice of the modified viscosity prescription. In Sect.~\ref{s.derivation} we derive the dispersion relation. In Sect.~\ref{s.long}
we present a detailed discussion of the long wavelength limit. In Sect.~\ref{s.solutions} we discuss solutions of the dispersion relation. Finally, in Sect.~\ref{s.summary} we summarize our results. Detailed discussion of their physical implications will be given in a separate paper.

%%%%%%%%%%%%%%%%%%%%%%%%%%%%%%%%%%%%%%%%%%%%%%%%%%%%%%%%%%%%%%%%%%%%%%%%%%%%%%
\section{Perturbative analysis of disk stability}
\label{s.dispersion}

We base our study on a linear perturbative analysis following the approach pioneered by
\cite{piran-78}. The unperturbed disk is assumed to be steady, so that time dependence can only be found in the perturbations. In the following, it is understood that the characteristic lengthscales of radial variation of unperturbed variables are $\sim r$. We consider small axisymmetric perturbations with radial wavenumber $k$, assuming that their wavelength $\lambda=2\pi/k$ satisfies the relation
\be
\label{ass.lambda}
H\ll\lambda \ll r,
\ee
where $H$ is the half-thickness of the disk.
(The assumption of geometrical thinness ceases to be valid for disks with
$L\gta \rm few \times 0.01\,L_{\rm Edd}$, so that strictly speaking our results apply only to the low luminosity region of the regime where the thermal-viscous instability would appear according to the standard model).
Since the radial flow is slow in comparison to the azimuthal motion, we make the usual assumption \citep{bb} in calculations of instability that $v_{r0}=0$, where the index $0$ will denote unperturbed quantities. We define the following dimensionless variables corresponding to the Eulerian perturbations of vertically integrated pressure ($P_1$), radial velocity ($v_{r1}$), surface density ($\Sigma_1$), disk thickness ($H_1$), and vertically integrated viscous stress ($T_1$),
\be \bar u = \frac{v_{r1}}{\Omega r}, ~~ \bar \sigma = \frac{\Sigma_1}{\Sigma_0},
~~ \bar \varpi = \frac{P_1}{P_0}, ~~  \bar h = \frac{H_1}{H_0}, ~~ \bar  \theta = \frac{T_1}{T_0},
\ee
where $\Omega=\sqrt{GM/r^3}$ is the Keplerian rotational frequency.
We assume that all of these quantitites represent complex waveforms $e^{n\Omega t-ikr}$, e.g., $\bar\varpi=\varpi e^{n\Omega t-ikr}$ and
\be
P(r,t)=P_0(r)+P_1(r,t)=P_0(r)\cdot(1+\varpi e^{n\Omega t-ikr}),
\ee
where $n$ stands for the dimensionless frequency, and $\varpi$ is a constant and uniform dimensionless amplitude. Negative values of the real part of the dimensionless frequency, $\Re (n)$, correspond to damped (stable) perturbations while positive values to exponentially growing (unstable) ones. The imaginary part of $n$ determines the frequency of the corresponding oscillations.

%-----------------------------------------------------------------------------
\subsection{Viscosity prescription}
\label{s.prescription}
In the standard approach, based on the $\alpha$-prescription \citep{shakura-73}, one assumes that the $r\varphi$ component of the stress tensor, $T_{r\varphi}$, is proportional to pressure:
\be
T_{r\varphi}(r,t)=-\alpha P(r,t),
\ee
where $T_{r\varphi}$ and $P$ are given in terms of the unperturbed and perturbed quantities as
\be T_{r\varphi} = T_0(1+\theta e^{n\Omega t-ikr}),
\label{e.trphi} \ee
\be P = P_0(1+\varpi  e^{n\Omega t-ikr}). \label{e.pressure}
\ee

In this work we assume a modified prescription for viscosity. Instead of assuming that $\alpha$ is constant both in time and radius we write
\be \alpha(r,t)=\alpha_0(1+\bar\alpha)=\alpha_0\left[1+(\theta e^{-n\Omega\tau}
-\varpi)e^{n\Omega t-ikr}\right], \label{e.alpha} \ee
with
\be T_0=-\alpha_0P_0. \label{e.alphazero}\ee
As 
\bea
&&-\alpha(r,t+\tau)P(r,t+\tau)=\nonumber\\
&&\quad = -\alpha_0(1+\theta e^{n\Omega t -ikr}
	-\varpi e^{n\Omega (t+\tau)-ikr})\times\nonumber\\
&&\quad\quad \times P_0(1+\varpi e^{n\Omega (t+\tau)-ikr})\nonumber\\
&&\quad \approx -\alpha_0 P_0(1+\theta e^{n\Omega t-ikr}),
\eea
taking into account Eqs.~\eqref{e.trphi} --- \eqref{e.alphazero} we have, through first order in the perturbation,
\be T_{r\varphi}(r,t)=-\alpha(r,t+\tau)P(r,t+\tau).\ee
Hence, our modified viscosity prescription, Eq.~\eqref{e.alpha}, corresponds to a stress which is still proportional to $\alpha P$ but is \textbf{advanced} with respect to this quantity by the time $\tau$, i.e., the value of stress at time $t$ is proportional to $\alpha P$ as measured at time $t+\tau$. In particular, a negative value of $\tau$ corresponds to a \textbf{delayed} stress: a perturbation of stress follows the perturbation in pressure after a delay of  $-\tau$.

We also introduce the heating to pressure response factor, defined as the ratio of the dimensionless amplitudes of stress and pressure perturbations,
\be\label{defksi}
\xi\equiv\frac\theta\varpi.
\ee
We assume $\xi$ to be real.
%-----------------------------------------------------------------------------
\subsection {Derivation of the dispersion relation}
\label{s.derivation}
In the following four subsections we derive perturbed forms of hydrostatic balance, mass conservation, angular momentum balance and energy equation for geometrically thin, axisymmetric accretion disks.

\subsubsection{Hydrostatic balance}

The balance of vertical forces, after integrating along the vertical coordinate, takes the following form,
\be \label{Hydro1} \Omega^2 H^2=\frac{P}{\Sigma}. \ee
Writing this form of the vertical force balance we assume the disk to be in hydrostatic equilibrium, since the thermal and secular timescales are much longer than the dynamical timescale.

Perturbing Eq.~\eqref{Hydro1} with small-amplitude axisymmetric perturbations,
and assuming that the azimuthal component of the velocity undergoes no change, we get
\be \label{Hydro2} \Omega^2 H_0^2 (1+\bar{h})^2=\frac{P_0
(1+\bar{\varpi})}{\Sigma_0 (1+\bar{\sigma})}. \ee
%and
%\be \label{Hydro3} \left(\Omega^2 H_0^2 \frac{\Sigma_0}{P_0}\right)
%(1+\bar{h})^2=\frac{(1+\bar{\varpi})}{ (1+\bar{\sigma})}. \ee
Using the unperturbed Eq.~\eqref{Hydro1}, we get
\be \label{Hydro4} (1+\bar{h})^2(1+\bar{\sigma})=1+\bar{\varpi}, \nonumber\ee
i.e., through linear order
\be \label{Hydro5} 1+2\bar{h}+\bar{\sigma}=1+\bar{\varpi}, \nonumber\ee
and finally
\be \label{Hydro8} 2\bar{h}= \bar{\varpi}-\bar{\sigma}. \ee

\subsubsection{Mass conservation}

The vertically integrated form of the continuity equation can be written as
\be \label{Mass1}\pder{\Sigma}{t} + \kpder{ }{r}{r}(r \Sigma v_r) =
0. \ee
Its perturbed form is
\be \label{Mass2}\pder{ }{t} \Sigma_0 (1+\bar{\sigma}) + \kpder{
}{r}{r}[r  \Sigma_0 (1+\bar{\sigma}) r \Omega \bar{u}] = 0. \ee
Neglecting terms of the second order ($\bar{u}\bar{\sigma}$)
and using Eq.~\eqref{Mass1} in its unperturbed form, $\partial{\Sigma_0}/\partial{t}=0$,
%\be \label{Mass3} \Sigma_0 \Omega n\bar{\sigma} +
%\frac{\Sigma_0 }{r}\pder{}r\left(r^2\Omega\bar u + r v_{r0}\bar\sigma\right) = 0. \ee
%Putting $v_{r0}=0$ and differentiating
we obtain
\be \label{Mass4} \Sigma_0 \Omega n\bar{\sigma} +
\Sigma_0\Omega \bar{u}\left[\pderln{(r^2 \Sigma_0\Omega)}{r}
-ikr \right] = 0. \ee
The logarithmic derivative term may be neglected on the strength of the assumption stated in Eq.~\eqref{ass.lambda}. Thus, we obtain the final relation
\be \label{Mass7} n\bar{\sigma} =ikr \bar{u}.\ee

\subsubsection{Angular momentum conservation}

The angular momentum conservation law is
\be \label{momentum}
 \Sigma v_r \frac1r\pder{ }{r} (r^2 \Omega) = -\kpder{ }{r^2}{r} (r^2 T_{r \varphi}).
\ee
For the first order perturbations we obtain
%\be \label{momentum2}  \Sigma_0 (1 + \bar{\sigma} )
%\Omega r\bar{u} \frac1r\der{}{r}(r^2 \Omega) = \frac{1}{r^2}\pder{}{r} \left(r^2 T_0(1+\bar\theta)\right)
% \ee
%Neglecting terms of higher order and taking into account Eq.~\eqref{momentum} written down for the unperturbed quantities, we get
\be
\Sigma_0\Omega r\bar u\frac1r\der{ }{r} (r^2 \Omega)=-\frac{1}{r^2}\pder{}{r} \left(r^2 T_0\bar\theta\right).
\ee
%Now we put $v_{r0}=0$, $T_0=-\alpha_0P_0$  and differentiate to obtain,
%\be
% \label{momentum3} \Sigma_0 \Omega \bar{u} r\left(2
%\Omega + r
%\der{\Omega}{r}\right) = - \frac{\alpha_0 P_0}{r} (2-ikr)\bar{\theta}.
% \ee
According to Eq.~\eqref{ass.lambda} the derivative on the right hand side is dominated by the $\partial\bar\theta/\partial r$ term. Using Eq.~(\ref{e.alphazero}), and introducing the speed of sound squared  $c^{2}_{s} \equiv P_0/\Sigma_0$, and
the vertical epicyclic frequency
\be
\kappa^2= \frac{2 \Omega}{r}\der{ \left(r^2\Omega \right)}{r},
\ee
we obtain
 \be \label{momentum5} \frac{\kappa^2 \bar{u}}{2 \Omega^2} -
ikr \alpha_0 \left( \frac{c_s}{r \Omega} \right)^2 \bar\theta  = 0, \ee
and finally,
\be \label{momentum6}\bar{u} = 2 ikr \alpha_0 \xi\left(
\frac{c_s}{r \kappa} \right)^2\bar{\varpi}. \ee
Incidentally, in our study $\kappa=\Omega$ and Eq.~\eqref{Hydro1} reads $c_s/(r\Omega)=H_0/r$, so $c_s/(r\kappa)=H_0/r$.

\subsubsection{Energy equation}

We start from the second law of thermodynamics in the form \citep[e.g.,][]{bb}
\be \begin{split}\pder{E}{t} + P \pder{\ln H}{ t} +
\kpder{}{r}{r}[r v_r(E+P)] -& v_r\pder{P}{r} +\\+ v_r P \pder{\ln
H}{r} &= Q_{vis}^{+} - Q_{rad}^{-},
\end{split} \ee
or simply,
\be \label{Ene1} \pder{E}{t} + \kpder{}{r}{r}(r v_rE)
+ \frac{P}{H} \left[\pder{H}{ t} + \kpder{}{r}{r}(r v_rH)\right]
 = Q_{vis}^{+} - Q_{rad}^{-},
 \ee
where E is the vertically integrated specific energy,
\be \label{Ene2} E = \left[3(1-\beta)+\frac{\beta}{\gamma-1}\right]
P \equiv A P, \ee
$\beta$ is the gas to total pressure ratio, $\gamma$ is the ratio of specific heats, $Q_{\rm vis}^+$ and $Q_{\rm rad}^-$ are the viscous heating and radiative cooling rates per unit area, respectively.

Let us differentiate both sides of Eq.~\eqref{Ene1} with respect to time. Assuming that $Q_{vis}^{+}$ is a
function of $P$ and $\alpha$ (following the $\alpha$-prescription) while $Q_{rad}^{-}$ is a function of $P$ and
$\Sigma$ (as is the case for radiative cooling in the optically thick regime), we obtain,

\bea \label{Ene3}
\pder{Q_{vis}^{+}}{t}&=&
 \left(\frac{\partial Q_{vis}^{+}}{\partial
P}\right)_{\Sigma, \alpha}\frac{\partial P}{\partial t} +
\left(\frac{\partial Q_{vis}^{+}}{\partial \alpha}\right)_{\Sigma,
P}\frac{\partial \alpha}{\partial t} \\
\pder{Q_{rad}^{-}}{t}&=&  \left(\frac{\partial
Q_{rad}^{-}}{\partial P}\right)_{\Sigma, \alpha}\frac{\partial
P}{\partial t} + \left(\frac{\partial Q_{rad}^{-}}{\partial
\Sigma}\right)_{P, \alpha}\frac{\partial \Sigma}{\partial t}
\eea
Introducing the perturbations and differentiating we get,
  \bea
\label{Ene4}\frac1{n\Omega} \pder{Q_{vis}^{+}}{t} &=&\left(\frac{\partial
Q_{vis}^{+}}{\partial P}\right)_{\Sigma, \alpha} P_0 \bar{\varpi} +
\left(\frac{\partial Q_{vis}^{+}}{\partial \alpha}\right)_{\Sigma,
P}\alpha_0
\bar{\alpha}\\
\frac1{n\Omega}\pder{Q_{rad}^{-}}{t} &=& \left(\frac{\partial
Q_{rad}^{-}}{\partial P}\right)_{\Sigma, \alpha}P_0 \bar{\varpi} +
\left(\frac{\partial Q_{rad}^{-}}{\partial \Sigma}\right)_{P,
\alpha}\Sigma_0\bar{\sigma}
  \eea

Finally, we have,
\be \label{Ene5}\pder{}t\left(Q_{vis}^{+}-Q_{rad}^{-}\right)=n\Omega^2 P_0\left(\bar{\varpi}G_{\varpi} +
\bar{\sigma}G_{\sigma}+\bar{\alpha} G_{\alpha}\right),
\ee
where
\bea
 \label{Ene6}
\quad\quad G_{\varpi} &=& \frac{1}{\Omega}
\left(\left(\pder{Q_{vis}^{+}}{P}\right)_{\Sigma, \alpha} -
\left(\pder{Q_{rad}^{-}}{P}\right)_{\Sigma,
\alpha}\right)  \\
\quad\quad G_{\sigma} &=& -\frac{1}{\Omega
c_s^2}\left(\pder{Q_{rad}^{-}}{\Sigma}\right)_{P,
\alpha}  \\
\quad\quad G_{\alpha} &=& \frac{\alpha_0}{\Omega P_0}
\left(\pder{Q_{vis}^{+}}{\alpha}\right)_{\Sigma, P}.
\eea
Following the $\alpha$-prescription, these quantities simplify to \citep{bb},
\bea \label{ggg1}
\quad\quad G_{\varpi} &=& \frac{3 \alpha_0 }{2}\cdot \frac{(2-5 \beta )}{ (4-3 \beta )}, \\\label{ggg2}
\quad\quad G_\sigma &=& \frac{3 \alpha_0 }{2}\cdot \frac{(2+3\beta)}{(4-3\beta)},\\\label{ggg3}
\quad\quad G_{\alpha} &=& \frac{3 \alpha_0 }{2}.
\eea

\hfill

The perturbation of $\beta$ is given by the following
expression \citep{cs},
\be \label{Ene13} \bar\beta\equiv\frac{\beta_1}{\beta_0} = \frac{1-\beta_0}{4-3\beta_0}(4
\bar{\sigma}-3 \bar{\varpi} - \bar{h}). \ee
Hence, like all the other time derivatives, the time derivative of $A$
is first order in the perturbation:
\be
\pder{A}{t}=\pder{}{t}\left[3(1-\beta)+\frac{\beta}{\gamma-1}\right]=
\beta_0\frac{4-3\gamma}{\gamma-1}\pder{\bar\beta}{t}.
\ee
The time derivative of the first term on the left hand side of Eq.~\eqref{Ene1} is given by,
\be
\npder{E}{t}{2}=\npder{ AP}{t}{2}  = P\npder{A}{t}{2} + 2
\pder{A}{t}\pder{P}{t} + A\npder{P}{t}{2},
\ee
where Eq.~\eqref{Ene2} has been taken into account. The derivatives
of $A$ and $P$ are both first order, so the middle term is second order
and may be neglected. The other terms are
\be \label{Ene11} A\npder P{t}{2} = A P_0 n^2\Omega^2\bar{\varpi} \ee
and
\be \label{Ene12} P\npder A{t}{2} = 
%P\npder{}{t}{2}\left[3(1-\beta)+\frac{\beta}{\gamma-1}\right] = 
P\frac{4-3\gamma}{\gamma-1}\npder\beta{t}{2} \ee
The second time derivative of $\beta=\beta_0(1+\bar\beta)$ equals,
\be \label{A} \npder\beta{t}{2} =n^2\Omega^2\beta_1, \ee
%= n^2 \Omega^2 \beta_0 \frac{1 -
%\beta_0}{4 - 3 \beta_0}(4 \bar{\sigma}-3 \bar{\varpi} - \bar{h})
so finally,
\be
\label{dedete}
\npder{E}{t}{2}=n\Omega^2 P_0\left(nA\bar\varpi+n\frac{4-3\gamma}{\gamma-1}\beta_0\bar\beta\right).
\ee

%Perturbing the time derivative of the second term in Eq.~\eqref{Ene1} and neglecting the higher order terms we obtain,
%\bea \quad\quad\nonumber\der{}{t}\left( P\pder{\ln H}{t}\right) &=& \der{}{t}\left(
%\frac{P_0(1+\bar\varpi)}{H_0(1+\bar h)}H_0n\Omega\bar{h}\right)\approx\\\label{Ene7}
%&\approx& n^2\Omega^2 P_0 \bar{h}.\eea

Perturbing the second term on the left hand side of Eq.~\eqref{Ene1},
keeping in mind the assumption of Eq.~\eqref{ass.lambda}, we get through first order
\be \label{lalala} \kpder{}{r}{r}(r^2 \Omega\bar{u}E)
=  - ikr \Omega A P_0 \bar{u}.
\ee
% \begin{split} \label{lalala} \kpder{}{r}{r}(r^2 \Omeg\bar{u}E)a=&\frac{A P_0 \bar{u}}{r} \pder{}{r}[r^2 \Omega (1 + \bar{\varpi})]= \\ 
%= & - ikr A P_0 \bar{u} \Omega. \end{split}\ee
The time derivative of this expression is
\be \label{Ene6b}\pder{}t\kpder{}{r}{r}(r^2 \Omega
\bar{u}E)=- ikrn \Omega^2 P_0 A\bar{u} \,.  \ee

The time derivatives of the terms involving $H$ in Eq.~\eqref{Ene1} are straightforward to compute. The first order terms are
\be \begin{split} 
\frac{P}{H} \left[\pder{H}{ t} + \kpder{}{r}{r}(r v_rH)\right] &=
P_0n\Omega\bar{h} + \frac{P_0}{H_0}\kpder{}{r}{r}(r^2\Omega\bar{u}H_0)=\\
&=P_0n\Omega\bar{h} -ikP_0r\Omega\bar{u}\end{split}
\ee
and their time derivative is
\be\label{Ene7}
\pder{}{t} \left[\frac{P}{H}\pder{H}{ t} + \frac{P}{H}\kpder{}{r}{r}(r v_rH)\right] =n\Omega^2P_0(n\bar{h} -ikr\bar{u}).
\ee
Collecting Eqs.~\eqref{Ene5}, \eqref{dedete}, \eqref{Ene6b} and \eqref{Ene7}, we obtain the final form of the perturbed energy equation,
\be
\label{Ene99901}
nA \bar{\varpi} + n\frac{4-3\gamma}{\gamma-1}\beta_0\bar\beta + n
\bar{h}-ikr \bar{u} (A+1)  = \bar{\varpi}G_{\varpi} +
\bar{\sigma}G_{\sigma}+\bar{\alpha}G_{\alpha}. \ee

\subsubsection{System of perturbed equations and the dispersion
relation}

Eqs.~\eqref{Hydro8}, \eqref{Mass7}, \eqref{momentum6} and \eqref{Ene99901} form a system of four algebraic equations.

The perturbation of $\alpha$, defined in Eq.~\eqref{e.alpha}, may be expressed in terms of $\bar\varpi$, and $\xi$ of Eq.~(\ref{defksi}):
\be
\frac{\bar{\alpha}}{\bar{\varpi}}\equiv \xi e^{-n\Omega\tau}-1.\ee
%where $\xi=\bar\theta/\bar\varpi$. 
Taking this relation into account we obtain four coupled homogeneous algebraic equations in the following form,

\be \label{syst}
 \left\{
\begin{array}{llll}
2\bar{h}= \bar{\varpi}-\bar{\sigma}, \\
\\
n\bar{\sigma}=ikr\bar{u},\\
\\
\bar{u} = 2 ikr \alpha_0 \left(
\frac{c_s}{r \kappa} \right)^2 \xi \bar{\varpi},\\
\\
A n{\bar{\varpi}} + n\frac{4-3\gamma}{\gamma-1}{\beta_0}\bar\beta  + n
{\bar{h}}-ikr {\bar{u}} (A+1)
= \\ ={\bar{\varpi}} G_{\varpi} + {\bar{\sigma}}G_{\sigma}+{\bar{\varpi}}(\xi
e^{-n\Omega\tau} -1)G_{\alpha},
\end{array} \right.
\ee
where $\bar\beta$ is given by Eq.~(\ref{Ene13}).
This system has nontrivial solutions for the four variables
$\bar{h}$, $\bar{\varpi}$, $\bar{\sigma}$, and $\bar{u}$, iff
\be \label{disperse} n^2 C_1 + n \left[ \xi FC_2 +G_{\alpha}( 1-\xi
e^{-n\Omega\tau}) - G_{\varpi}\right]  + \xi F G_{\sigma}= 0,\ee
where 
\bea
\quad\quad F &\equiv& 2 \alpha_0 \left( \frac{k c_s}{\kappa}\right)^2,\\
\quad\quad C_1 &\equiv& A +\frac{7}{2}B + \frac{1}{2}, \\
\quad\quad C_2 &\equiv& A + \frac{9}{2} B +\frac{3}{2} , \\
\quad\quad B &\equiv& \beta \frac{(3 \gamma -4)(1 - \beta)}{(\gamma -
1)(4 -3\beta)},
\eea
and $A$ is given in Eq.~\eqref{Ene2}.
For $\beta=0$ the coefficients of Eq.~\eqref{disperse} do not depend on $\gamma$.
For $\xi=1$ and $\Omega\tau = 0$ the condition given in Eq.~\eqref{disperse} simplifies to the standard dispersion relation for perturbed accretion disks \citep[e.g.,][]{bb}.  Note that $0\le \beta \le 1$, and that for
$\gamma \ge 4/3$, $A$ and $B$ satisfy $A>0$, and $B\ge 0$. Thus,  both $C_1$ and $C_2$ are positive for any value of $\beta$. In the following, we will take
$\gamma=5/3$ whenever a specific value is required for numerical results.
We also specialize to $F=2\alpha_0(kH_0)^2$, in accordance with the comment following Eq.~\eqref{momentum6}.

%%%%%%%%%%%%%%%%%%%%%%%%%%%%%%%%%%%%%%%%%%%%%%%%%%%%%%%%%%%%%%%%%%%%%%%%%%%%%%
\section{Long-wave limit}
\label{s.long}

%\section{Long-wave limit}
%\label{s.solutions}
Let us first consider the solutions of the dispersion relation Eq.~(\ref{disperse})
in the long wavelength limit, i.e., $kH \rightarrow 0$, 
which
is very useful in classifying solutions of arbitrary wavelength.
In the limit $kH \rightarrow 0$ we neglect terms proportional to $F$ and obtain
\be 
\label{lw1}n\left[n C_1  +  (1-\xi
e^{-n\Omega\tau})G_{\alpha}-G_{\varpi}\right] = 0, 
\ee
with a trivial solution $n=0$. Dividing by $n\neq 0$ and using Eqs.~\eqref{ggg1} - \eqref{ggg3} we get for the remaining solutions
\be \label{lw00}n C_1  + \frac{3}{2} \alpha_0 \left(1-\xi
e^{-n\Omega\tau}-f\right) = 0,\ee
where
\be
f\equiv \frac{G_{\varpi}}{G_{\alpha}}=\frac{2-5 \beta}{4-3 \beta}.
\ee

%-------------------------------------------------------------------------
\subsection{The case of no delay, {$\Omega\tau=0$}}
Solving Eq.~\eqref{lw00} is easy for $\Omega\tau=0$. In this case $n$ is real and is given by,
\be
n_0=\frac{3\alpha_0(\xi+f-1)}{2C_1}.
\label{eee.6}
\ee
As $C_1$ is positive for all values of $\beta$, the sign of $n$ is determined by the sign of $\xi+f-1$. For $\xi=1$, $n_0=G_{\varpi} /C_1$ and  it is positive for $\beta<2/5$ and negative for $\beta>2/5$. The latter inequality is the standard condition for disk stability.

\begin{figure}
\centering
 \includegraphics[width=.95\columnwidth]{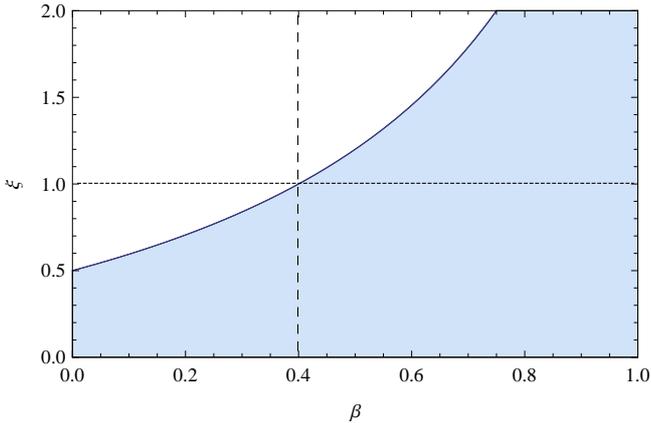}
\caption {The shaded area denotes the region in the ($\beta$, $\xi$) plane for which the real root of Eq.~\eqref{lw00} is negative for all $\Omega\tau\ge0$.}
 \label{f.tauvsxi}
\end{figure}

In general, we get the following criterion for the negative sign of $n$,
\be \label{lw05} \xi < 1-f=\frac{2(1+\beta)}{4-3\beta}. \ee
The shaded area on Fig.~\ref{f.tauvsxi} presents the region in the $(\beta,\xi)$ plane for which $n$ is negative. Note that it suffices to decrease the amplitude of stress variations by a factor of two ($\xi=0.5$) to stabilize the disk for all $\beta$.
Despite the fact that the condition \eqref{lw05} has been derived assuming $\Omega\tau=0$, it remains satisfied for the negative roots of Eq.~\eqref{lw00} for all $\Omega\tau \ge0$, as will be shown in Section~\ref{s.realroot}.

%++++++++++++++++++++++++++++++++++++++++++++++++++++++++++++++++
% stability regions on (beta, xi)
%+++++++++++++++++++++++++++++++++++++++++++++++++++++++++++++++++
%++++++++++++++++++++++++++++++++++++++++++++++++++++++++++++++++

%----------------------------------------------------------------------------
\subsection{{Arbitrary $\Omega\tau$}}

Let us now consider the general case of $\Omega\tau\neq 0$. Equation~\eqref{lw00} is no longer trivial as it involves an exponential function of $n$. An infinite number of complex solutions is expected as the exponential function $e^{-n\Omega\tau}$ involves periodic trigonometric functions whenever the imaginary part of $n$ is nonzero, $\Im(n)\ne0$. All imaginary solutions are conjugate, i.e., if $x$ satisfies Eq.~\eqref{lw00} then $x^*$ is also a solution. As will be shown in the following section, no more than two real solutions may exist. To find the roots of a nonlinear complex equation such as Eq.~\eqref{lw00} one has to use numerical methods.
We used the \texttt{MINPACK} routines \citep{minpacka}. First, the locations of minima of the absolute value of the left hand side of Eq.~\eqref{lw00} were roughly estimated. The values so obtained served as starting points for the nonlinear solver.

\begin{figure*}
\centering
 \subfigure {\includegraphics[height=.33\textwidth]{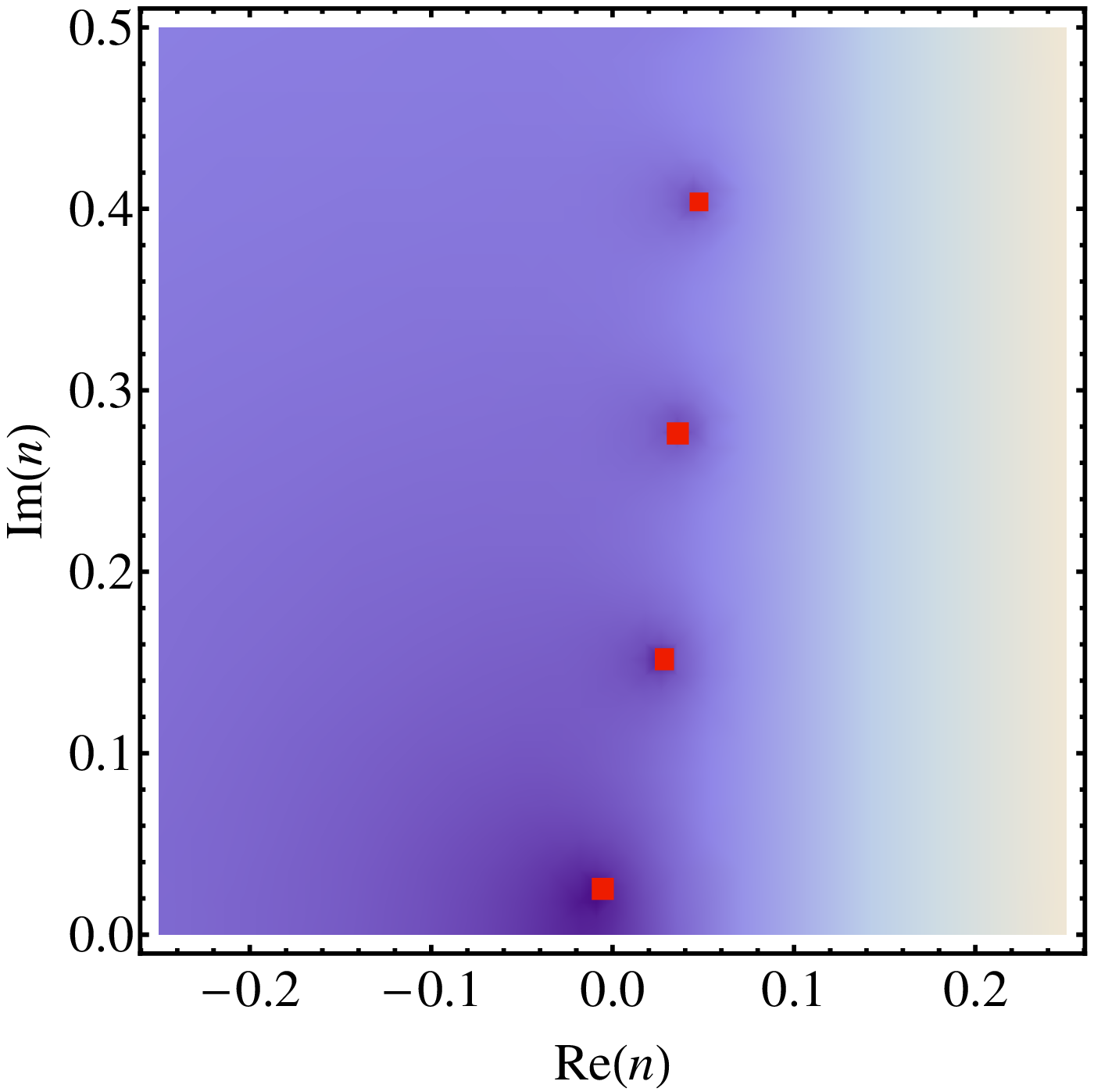}}
 \subfigure {\includegraphics[height=.33\textwidth]{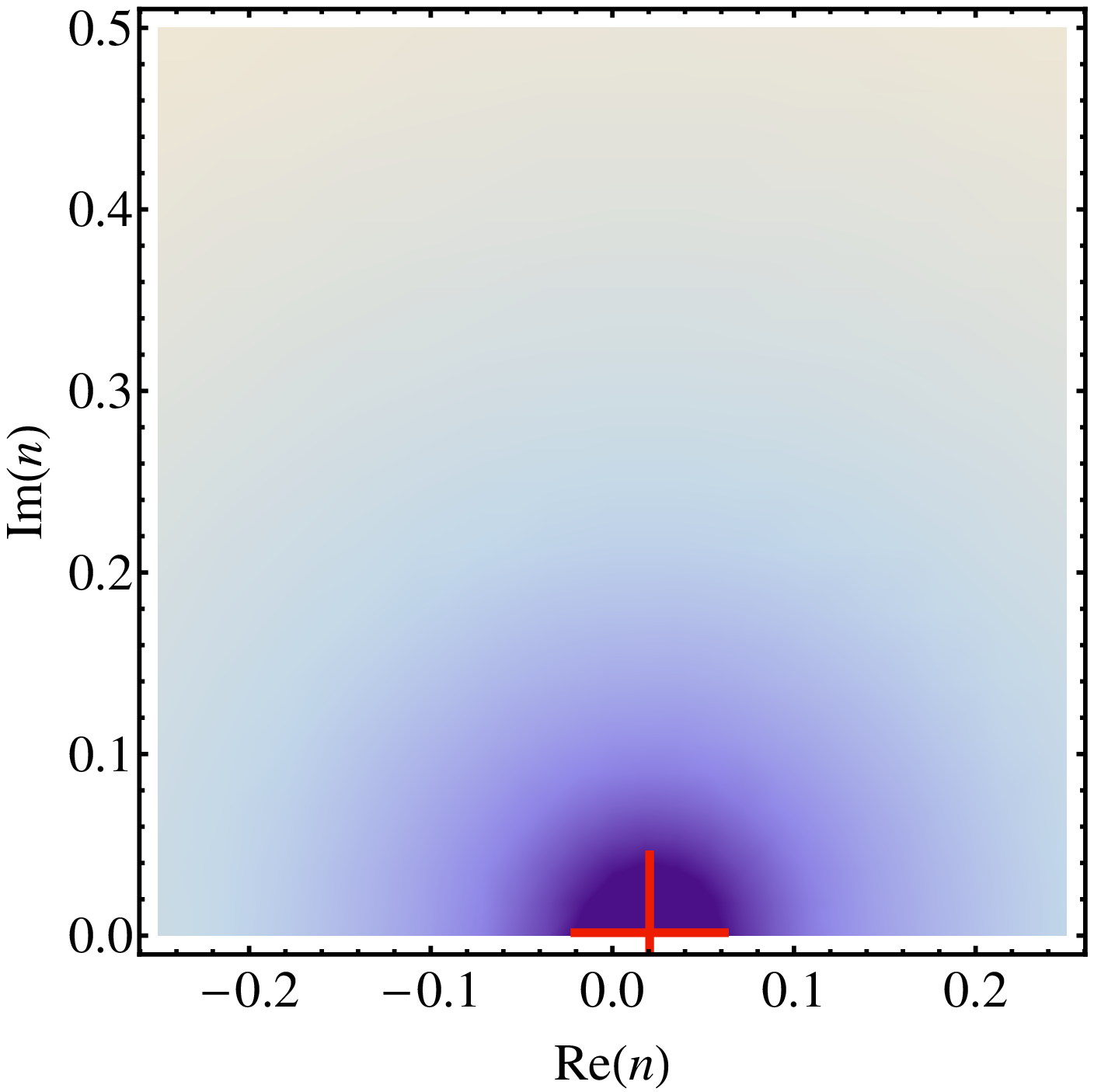}}
 \subfigure {\includegraphics[height=.33\textwidth]{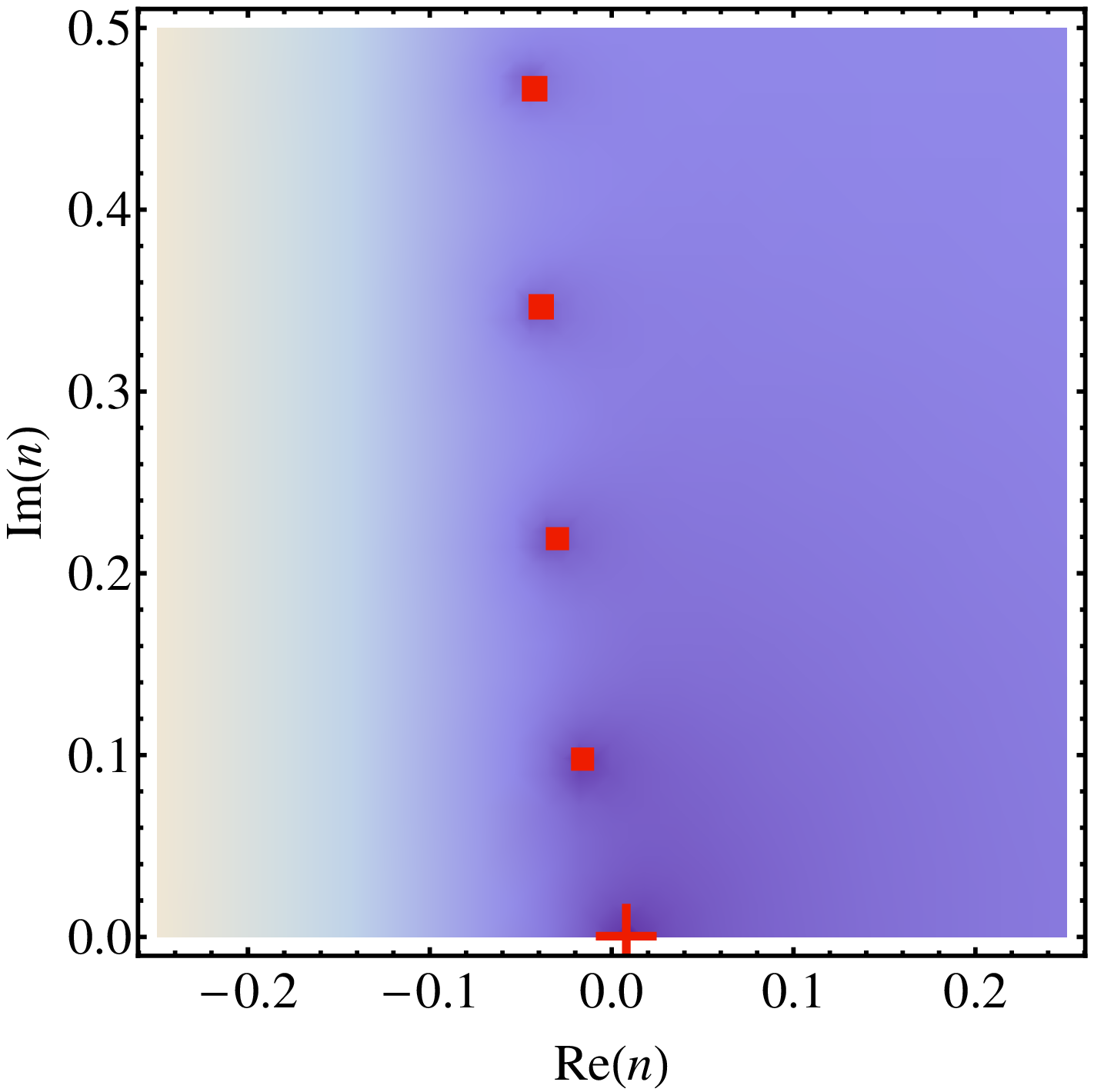}}
\caption {Maps presenting absolute values of the left hand side of
  Eq.~\ref{lw00} (dispersion relation in the limit of long
  waves) on the complex plane for $\beta=0$, $\xi=1$, $\alpha=0.1$ and three values
  of $\Omega\tau=-50$ (left), $0$ (middle) and $50$ (right
  panel). Crosses and rectangles denote locations of the real and complex solutions, respectively.}
 \label{f.colormaps}
\end{figure*}

In Fig.~\ref{f.colormaps} we plot color-coded absolute values of the left hand side of Eq.~\eqref{lw00} for $\beta=0$, $\xi=1$, $\alpha=0.1$ and three values of $\Omega\tau=-50$ (left), $\Omega\tau=0$ (middle), $\Omega\tau=50$ (right panel). Only $\Im(n)\ge0$ regions are shown. The darker the color, the smaller the value. Red crosses denote real solutions while red squares show locations of solutions with non-zero imaginary part. For each of the values $\Omega\tau=0$ and $\Omega\tau=50$ a single real solution exists, and it satisfies $\Re(n)>0$. For {$\Omega\tau\neq 0$} there is an infinite number of complex solutions. The sign of their real part (with the exception of the first imaginary root when $\Omega\tau<0$) is in general opposite to the sign of $\Omega\tau$.

%++++++++++++++++++++++++++++++++++++++++++++++++++++++++++++++++
% Figure: color maps
%+++++++++++++++++++++++++++++++++++++++++++++++++++++++++++++++++
%++++++++++++++++++++++++++++++++++++++++++++++++++++++++++++++++

\begin{figure}
\centering
 \includegraphics[width=.95\columnwidth]{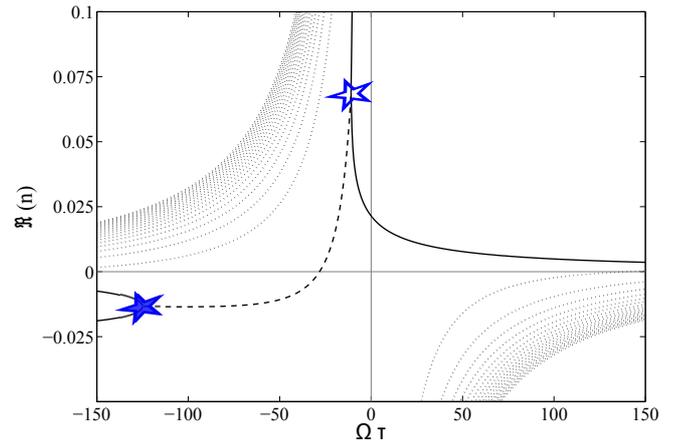}
\caption {Real part of solutions of
  Eq.~\eqref{lw00} for $\beta=0$, $\xi=1$ and $\alpha=0.1$
as a function of
  $\tau$.  
%The trivial solution, $n(\Omega\tau)\equiv0$, has been omitted for clarity.
The solid lines denote real
solutions, the dashed line shows complex solutions linking the real branches, while the dotted lines
present periodic complex solutions (only first $20$ solutions are drawn). Stars denote points where the real branches merge into the complex one (and its conjugate).}
 \label{f.f00}
\end{figure}

Fig.~\ref{f.f00} presents the real part of solutions of Eq.~\eqref{lw00} in the ($\Omega\tau$,$\Re(n)$) plane for the chosen values of $\alpha$, $\beta$ and $\xi$. Solid lines present real solutions (crosses in Fig.~\ref{f.colormaps}). Dotted lines correspond to ordinary complex roots (squares in Fig.~\ref{f.colormaps})---only the first 20 are plotted. The dashed line connecting two real branches is a special class of complex solutions.

Taking $\Omega\tau=0$ we recover the real solution given by Eq.~(\ref{eee.6}). For small but negative $\Omega\tau$ there are two real solutions, one of which satisfies $\lim_{\Omega\tau\to0^-}\Re (n)=+\infty$. These two real solutions converge to each other and merge into complex conjugate solutions (empty star in Fig.~\ref{f.f00}) at a critical value of $\Omega\tau_1\approx -11$ . These roots split into two real solutions, but this time negative, at $\Omega\tau_2\approx-125$ (solid star). The upper branch of these two approaches $\Re(n)=0^-$ for $\Omega\tau\to -\infty$. For $\Omega\tau>0$ there is only one real solution, which is positive and approaches $\Re(n)=0^+$ for $\Omega\tau\to +\infty$.

In addition to these solutions, an infinite number of periodic complex solutions exists (dotted lines). For most of the range of $\Omega\tau$ presented in Fig.~\ref{f.f00} they appear in the second and fourth quadrant of ($\Omega\tau,\Re(n)$) plane for negative and positive $\Omega\tau$, respectively. For $|\Omega\tau|\gtrsim 150$, however, the first complex root, and subsequently the others too (but at much larger absolute values of $\Omega\tau$), crosses the $\Re (n)=0$ axis.

%++++++++++++++++++++++++++++++++++++++++++++++++++++++++++++++++
% ksi=1, beta=0 vs tau & alpha
%+++++++++++++++++++++++++++++++++++++++++++++++++++++++++++++++++
%+++++++++++++++++++++++++++++++++++++++++++++++++++++++++++++++++

%----------------------------------------------------------------------------
\subsection{Real roots}
\label{s.realroot}

Let us examine the real branches in detail. Eq.~\eqref{lw00} may be rewritten in the following form,
\be
an+b=ce^{-n\Omega\tau},
\label{eee.1}
\ee
where
\bea
\quad\quad a&=&C_1>0,\\
\quad\quad b&=&\frac32\alpha_0(1-f)>0,\\
\quad\quad c&=&\frac32\alpha_0\xi>0.
\eea
Convexity
of the exponential function on the right hand side of Eq.~\eqref{eee.1} implies that there are no more than two real numbers $n$ satisfying this equation.
%It is clear that for real $n$ the linear function on the left hand side of Eq.~\eqref{eee.1} may have no more than two intersections with the exponential function on the right hand side. 
For $\Omega\tau\ge0$ the exponential function is decreasing with $n$ and therefore has exactly one intersection with $an+b$, and only for $n>0$ (or $n=0$, or $n<0$) as long as $b/c<1$ (or $b/c=1$, or $b/c>1$, respectively). This yields the same criterion for stability as in the $\Omega\tau=0$ case, inequality \eqref{lw05}. In this sense, advanced heating [$T_{r\phi}(t)=-\alpha P(t+\tau)$ with $\tau>0$] does not alter the viscous and thermal stability properties of the disk. 

For $\Omega\tau<0$ the exponential function is, on the contrary, increasing and may not intersect the linear one at all, may have a single intersection point or may cross it twice. All three cases occur and are clearly visible in Fig.~\ref{f.f00}, the single solutions are marked by stars and the complex branch connecting these real solutions corresponds to the no intersection case.

\begin{figure}
\centering
 \includegraphics[width=.95\columnwidth]{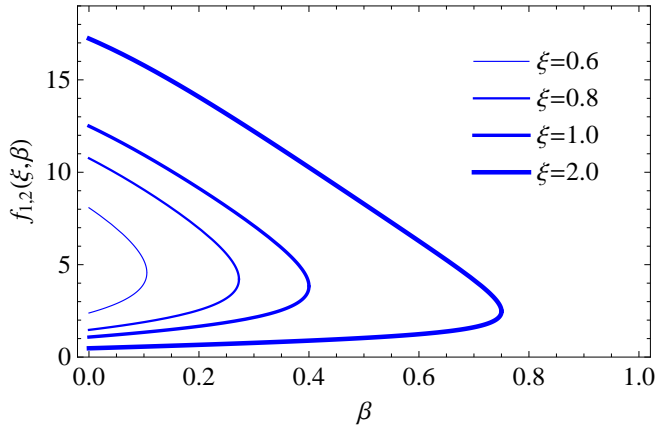}
\caption {$f_{1,2}(\xi,\beta)$ (Eq.~\ref{eee.5}) dependence on $\beta$ for various $\xi$. In this figure, we take $\gamma=5/3$.
}
 \label{f.omegataun}
\end{figure}

The condition for a single real solution of Eq.~\eqref{lw00} for $\Omega\tau<0$  may be obtained by matching the gradient of the linear function with the
derivative of the right hand side of Eq.~\eqref{eee.1},
\be
\label{eee.2}
a=\der{}{n}\left(ce^{-n\Omega\tau}\right)=-c\Omega\tau e^{-n\Omega\tau}.
\ee
Eqs.~\eqref{eee.1} and \eqref{eee.2} form a set of two equations corresponding to the linear function being tangent to the exponential one at a given $n$ and providing the condition, on $\Omega\tau$ for instance, under which Eq.~\eqref{lw00} has only  one real solution. After some algebra we get,
\be
-\frac{b}{ce}=-\frac{1-f}{\xi e}=we^w,
\label{eee.3}
\ee
where
\be
w\equiv\frac{b}{a}\Omega\tau=\frac{3\alpha_0(1-f)}{2C_1}\Omega\tau.
\ee
{Solutions
of Eq.~\eqref{eee.3} are given by the multivalued Lambert W function {\citep{wright59}} of its
left hand side. Real solutions exist only for values of the left hand side of the equation
greater or equal than $-1/e$. This condition corresponds to,}
\be \xi \ge \frac{2(1+\beta)}{4-3\beta} \label{eee.lw00bisbis}.\ee
As $we^w$ has a single minimum, of value $-1/e$ at $w=-1$, Eq.~\eqref{eee.3}
has two solutions for all $-1<w<0$, i.e., when the inequality in (\ref{eee.lw00bisbis}) is sharp.
 As long as condition \eqref{eee.lw00bisbis} is satisfied, the values of $\Omega\tau$ for which Eq.~\eqref{lw00} has only one real solution are therefore given by
\be
\Omega\tau_{1,2}=-f_{1,2}(\xi,\beta)\frac1{\alpha_0},
\label{eee.5}
\ee
where  $f_{1,2}(\xi,\beta)$ is double-valued. We take $f_1\le f_2$, i.e., $\Omega\tau_2\le\Omega\tau_1<0$. In Fig.~\ref{f.omegataun} we plot values of $f_{1,2}(\xi,\beta)$ as a function of $\beta$ for various values of $\xi$. For the standard case of $\xi=1$ and radiation pressure dominated disks ($\beta=0$) we have $f_1(1,0)\approx1.08$ and $f_2(1,0)\approx12.50$. Thus, for, e.g., $\alpha=0.1$, Eq.~\eqref{lw00} has only one real solution for $\Omega\tau_1\approx-10.8$ and $\Omega\tau_2\approx-125.0$ (corresponding to the stars in Fig.~\ref{f.f00}).
A moment's reflection leads to the conclusion that there are two positive roots of Eq~\eqref{lw00} for $\Omega\tau_1<\Omega\tau<0$, two negative ones for $\Omega\tau<\Omega_2\tau$, and no real roots for $\Omega_2\tau<\Omega\tau<\Omega\tau_1$. When condition \eqref{lw05}is satisfied, instead of \eqref{eee.lw00bisbis}, there are two solutions for $n$, one positive and one negative.
%++++++++++++++++++++++++++++++++++++++++++++++++++++++++++++++++
% stability regions on (beta, xi)
%+++++++++++++++++++++++++++++++++++++++++++++++++++++++++++++++++
%++++++++++++++++++++++++++++++++++++++++++++++++++++++++++++++++

%-------------------------------------------------------------------------------
\subsection{Limits of infinite time offset ($\Omega\tau\to\pm\infty$)}

Let us now investigate the limits of real solutions of Eq.~\eqref{lw00}. Assuming that $n$ is finite, Eq.~\eqref{eee.1} simplifies in the limit of $n\Omega\tau\to+\infty$ to,
\be
n_{-}=-\frac{b}{a}=-\frac32\frac{\alpha_0}{C_1}\frac{2(1+\beta)}{4-3\beta}<0.
\label{eee.nminus}
\ee
Therefore, this value is valid only for $\Omega\tau\to-\infty$. For the standard choice of $\alpha=0.1$, $\beta=0$ and $\xi=1$ one gets,
\be
n_{-}=-0.021,
\ee
which corresponds to the limit of the lower real branch in the third quadrant of Fig.~\ref{f.f00}.

To find limits of the other two branches let us assume that $n\Omega\tau$ remains finite for $|\Omega\tau|\to\infty$. This assumption implies $n\to0$ and Eq.~\eqref{eee.1} takes in this limit the following form,
\be
b=c e^{-n\Omega\tau}.
\ee
Therefore, $n$ has to fulfill the relation,
\be
n=-\frac1{\Omega\tau}\log \frac bc.
\ee
The sign of the logarithm depends on $\xi$ and $\beta$:
\bea
\quad\quad\log b/c < 0 &{\rm for} & \xi>\frac{2(1+\beta)}{4-3\beta},\label{eee.10}\\
\quad\quad\log b/c > 0 &{\rm for} & \xi<\frac{2(1+\beta)}{4-3\beta}.\label{eee.11}
\eea
Thus, for $\xi$ and $\beta$ satisfying \eqref{eee.10} $n$ approaches $0^+$ for $\Omega\tau\to+\infty$ and $0^-$ for $\Omega\tau\to-\infty$. If condition \eqref{eee.11} is satisfied we have $\lim_{\Omega\tau\to+\infty}n=0^-$ and $\lim_{\Omega\tau\to-\infty}n=0^+$.
%In the following paragraphs we study the impact of $\alpha_0$, $\beta$ and $\xi$.

\hfill

\subsection{{Parameter study}}
\label{s.study}
\begin{figure}
\centering
 \includegraphics[width=.95\columnwidth]{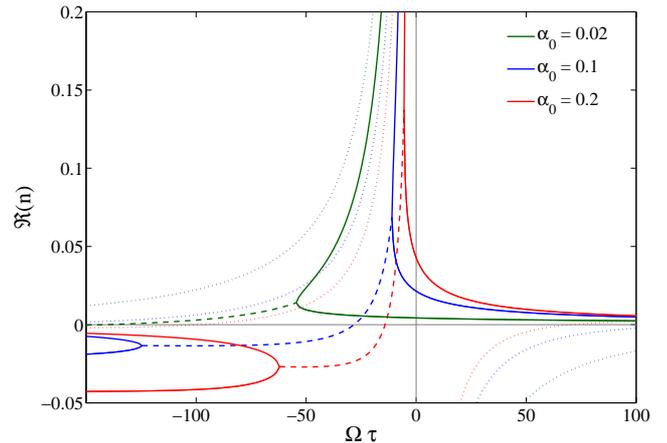}
\caption {Real part of solutions of
  Eq.~\eqref{lw00} for $\beta=0$ and $\xi=1$ as a function of
  $\tau$. Solutions for three values of $\alpha$ are presented with different colors (increasing values of $\alpha_0$ from left to right). Solid lines are for real
solutions while dotted and dashed lines show the real part of complex solutions. Only the first periodic exponential root is plotted. }
 \label{f.ksi1beta0}
\end{figure}

In Fig.~\ref{f.ksi1beta0} we show the roots of Eq.~\eqref{lw00} for $\beta=0$, $\xi=1$ and three values of $\alpha_0=0.02$ (green), $\alpha_0=0.1$ (blue) and $\alpha_0=0.2$ (red line). The second case corresponds to Fig.~\ref{f.f00}. For clarity, from among the infinite number of complex periodic solutions (dotted lines), only the first one {(in the sense of the smallest modulus value)} is plotted. All the curves have qualitatively similar shapes---the sign of solutions in a given region does not depend on $\alpha_0$ (Table~\ref{ttt.1}). However, the values of $\Omega\tau_1$ and $\Omega\tau_2$, which limit the regions with double real solutions for $\Omega\tau<0$, are sensitive to the value of $\alpha_0$. Eq.~\eqref{eee.5} predicts that their values are inversely proportional to $\alpha_0$ and therefore, e.g., the region with two negative roots of  Eq.~\eqref{lw00} extends to larger values of $\Omega\tau$ (closer to $\Omega\tau=0$) for high values of $\alpha$. The limit for the lower real branch ($n_-$) also depends on $\alpha_0$ according to Eq.~\eqref{eee.nminus}---the higher the value of $\alpha_0$, the lower the value of $n_-$.	

%++++++++++++++++++++++++++++++++++++++++++++++++++++++++++++++++
% ksi=1, beta=0 vs tau & alpha
%+++++++++++++++++++++++++++++++++++++++++++++++++++++++++++++++++

\begin{figure}
\centering
 \includegraphics[width=.95\columnwidth]{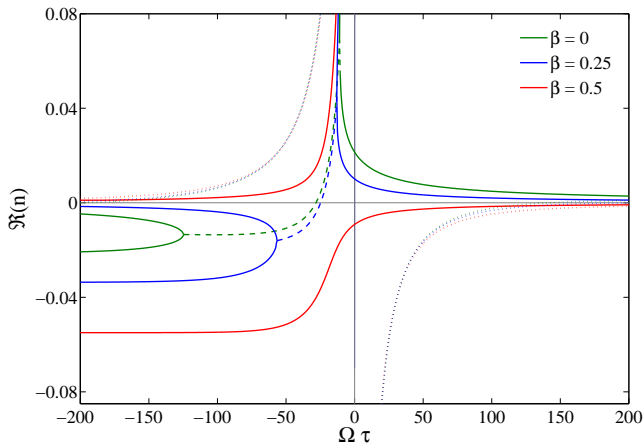}
\caption {Real part of solutions of
  Eq.~\eqref{lw00} for $\alpha=0.1$ and $\xi=1$ versus $\tau$. Solutions for
different values of $\beta$ are presented (for the solid lines in the $\Omega\tau>0$ region, $\beta$ increases from top to bottom). Solid lines are for real
solutions while dotted and dashed lines show the real part of complex solutions. Only the first periodic exponential root is plotted. In this figure, we take $\gamma=5/3$.}
 \label{f.ksi1alpha01}
\end{figure}

The impact of $\beta$ on solutions of the dispersion relation for $\alpha=0.1$ and $\xi=1$ is presented in Fig.~\ref{f.ksi1alpha01}. For $\xi=1$ the crucial inequality \eqref{lw05} corresponds to $\beta>2/5$. For values of $\beta$ smaller than this critical value solutions exhibit qualitatively the same behavior as discussed previously. The values of $\Omega\tau_{1,2}$ and $n_-$ depend on $\beta$ according to Eqs.~\eqref{eee.5} and \eqref{eee.nminus}, respectively. Once $\beta$ exceeds $2/5$ the character of the solution changes. For $\Omega\tau\ge 0$ the root is negative and approaches $0^-$ with $\Omega\tau\to+\infty$. For any negative value of $\Omega\tau$ there are two real solutions with opposite signs---the complex conjugate branch connecting the real solutions does not appear and $\Omega\tau_{1,2}$ are not defined.

%++++++++++++++++++++++++++++++++++++++++++++++++++++++++++++++++
% ksi=1, alpha=0.1 vs beta & tau
%+++++++++++++++++++++++++++++++++++++++++++++++++++++++++++++++++
%++++++++++++++++++++++++++++++++++++++++++++++++++++++++++++++++

\begin{figure}
\centering
 \includegraphics[width=.95\columnwidth]{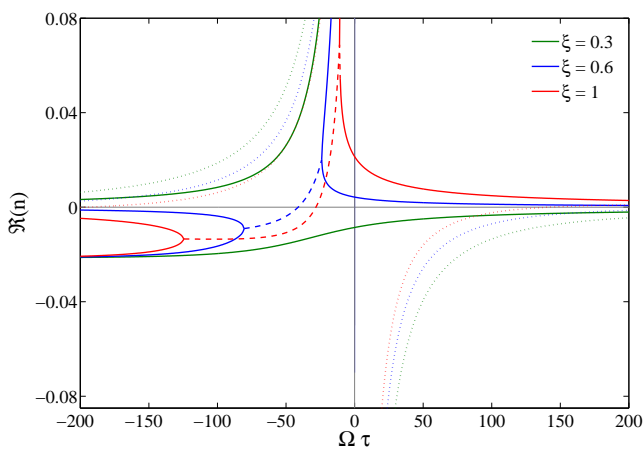}
\caption {Same as Fig.~\ref{f.ksi1alpha01} but for $\alpha=0.1$,
  $\beta=0$ and varying $\xi$.  For the solid lines in the $\Omega\tau>0$ region, $\xi$ decreases from top to bottom.}
 \label{f.alpha01beta0}
\end{figure}

Very similar behavior is shown in Fig.~\ref{f.alpha01beta0} which presents the impact of $\xi$ for $\alpha=0.1$ and $\beta=0$. For this case inequality \eqref{lw05} is not satisfied for $\xi>1/2$. The solutions change their nature once $\xi$ goes below this value, similarly to solutions with $\beta<2/5$ discussed in the previous paragraph. In accordance with Eq.~\eqref{eee.nminus}, $n_-$, the limit of the lower real branch at $\Omega\tau\to-\infty$,  does not depend on $\xi$.

%++++++++++++++++++++++++++++++++++++++++++++++++++++++++++++++++
% beta=0, alpha=0.1 vs xi & tau
%+++++++++++++++++++++++++++++++++++++++++++++++++++++++++++++++++
%++++++++++++++++++++++++++++++++++++++++++++++++++++++++++++++++

In Table~\ref{ttt.1} we summarize general features of the real solutions of Eq.~\eqref{lw00} which have been derived in this section.

\begin{table*}
%\newcolumntype{C}{>{\centering\arraybackslash}}%
\centering
\begin{tabular}{lccccc}
%\caption{ab cd}
\hline \hline
    & $\Omega\tau\to-\infty$& $\Omega\tau<0$ & $\Omega\tau=0$ &$\Omega\tau>0$ & $\Omega\tau\to+\infty$ \\
\hline
\begin{large}
$\xi>\frac{2(1+\beta)}{4-3\beta}$ \end{large}&
\begin{minipage}{.15\linewidth}
\begin{center}
$n_1\to n_-<0$\\
$n_2\to0^-$
\vskip1pt
\end{center}
\end{minipage}
&

\begin{minipage}{.35\linewidth}
\begin{center}
\hfill\\
two negative roots for $\Omega\tau<\Omega\tau_2$\\
complex conjugate roots for $\Omega\tau_2<\Omega\tau<\Omega\tau_1$\\
two positive roots for $\Omega\tau_1<\Omega\tau$\\
\vspace{.13cm}
\end{center}
\end{minipage}

& $n=n_0>0$& $n>0$ & $n\to0^+$ \\
\hline
\begin{large}
$\xi<\frac{2(1+\beta)}{4-3\beta}$ \end{large}&
\begin{minipage}{.15\linewidth}
\begin{center}
\hfill\\
$n_1\to n_-<0$\\
$n_2\to0^+$\\
\vspace{.13cm}
\end{center}
\end{minipage}
& two roots with opposite signs& $n=n_0<0$& $n<0$ & $n\to0^-$ \\
\hline \hline
\end{tabular}
\caption{The long wavelength limit: general characteristic of real solutions of Eq.~\eqref{lw00}. The quantities ~$n_0$, $\Omega\tau_{1,2}$ and $n_-$ are given by Eqs.~\eqref{eee.6}, \eqref{eee.5} and \eqref{eee.nminus}, respectively.}
\label{ttt.1}
\end{table*}

%%%%%%%%%%%%%%%%%%%%%%%%%%%%%%%%%%%%%%%%%%%%%%%%%%%%%%%%%%%%%%%%%%%%%%%%%%%%%%
\section{Solutions of the dispersion relation}
\label{s.solutions}
\subsection{Short-wave limit}
\label{s.szort}

Let us now consider the short-wave limit ($kH\to\infty$) of Eq.~\eqref{disperse}. Strictly speaking, this limit violates the first of the assumptions of Eq.~\eqref{ass.lambda}.
If we assume that $n$ is finite then the terms with $F$ dominate,

\be \label{szx} n \xi FC_2 + \xi F G_{\sigma}= 0. \ee
Despite the fact that $n$ in general is complex, in this case it must be real to satisfy this real equation. We obtain,
\be \label{szcx2} n = -\frac{G_{\sigma}}{C_2}. \ee
Both $G_\sigma =  (3 \alpha_0 /2){(2+3\beta)}/{(4-3\beta)}$ and $C_2$ are positive. Thus, this value of $n$ is negative and satisfies the dispersion relation in the short-wave limit for all values of parameters $\xi$ and $\Omega\tau$.

Let us now assume that the absolute value of $n$ is large ($|n|\gg1$ and $|n|\gg G_\sigma/C_2$). In this (and $kH\gg1$) limit Eq.~\eqref{disperse} reduces to
\be
\label{szsz.1}
nC_1+\xi FC_2-G_\alpha\xi e^{-n\Omega\tau}=0.
\ee
Assuming in addition that $n\Omega\tau\ge0$ we get
\be
nC_1+\xi FC_2=0,
\ee
which is satisfied for
\be
n=-\frac{\xi F C_2}{C_1}\to -\infty.
\label{szsz.2}
\ee
Hence, this limit is valid for $\Omega\tau\le0$.

If, on the contrary, we assume $n\Omega\tau<0$ then Eq.~\eqref{szsz.1} reduces to,
\be
\xi FC_2-G_\alpha\xi e^{-n\Omega\tau}=0.
\ee
The solution of which is,
\be
n=-\frac1{\Omega\tau}\log\frac{FC_2}{G_\alpha}.
\ee
Thus, we have two additional limits,
\bea
\quad\quad n\to -\infty & {\rm for} & \Omega\tau>0, \label{sz.10}\\
\quad\quad n\to +\infty & {\rm for} & \Omega\tau<0. \label{sz.20}
\eea

Summing up, there is a common limit $n=-G_\sigma/C_2$ for all values of $\Omega\tau$. In addition, solutions with $\Omega\tau\ge 0$ have another branch approaching $-\infty$ while for negative $\Omega\tau$ two branches are expected approaching both $+\infty$ and $-\infty$.

%----------------------------------------------------------------------------
\subsection{{Arbitrary wavelength}}

In this section we consider solutions of the dispersion relation for an arbitrary value of the wavelength $\lambda=2\pi/k$. 

\begin{figure}
\centering
 \includegraphics[width=.95\columnwidth]{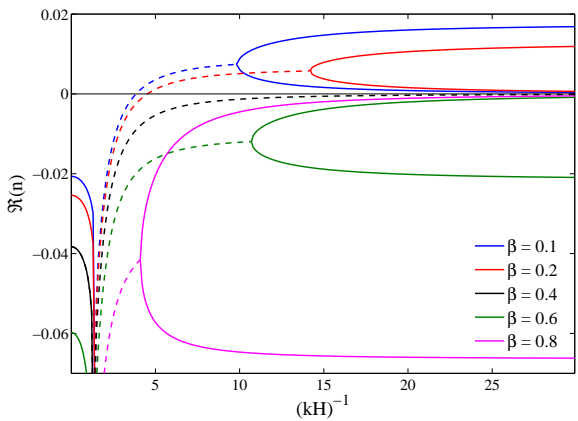}
\caption {Real part of solutions of Eq.~\eqref{disperse} as a
  function of the wave length $1/(kH)$ for $\alpha_0=0.1$, $\xi=1$, $\Omega\tau=0$ and
  various values of $\beta$. The line convention is the same as in previous figures. In this figure, we take $\gamma=5/3$.}
 \label{f.khrtau0}
\end{figure}

We start with discussing the standard case with no time lag ($\Omega\tau=0$) and $\xi=1$. Fig.~\ref{f.khrtau0} presents the real part of solutions of Eq.~\eqref{disperse} obtained assuming $\alpha=0.1$ and various values of $\beta$. 
The limit of $1/(kH) \to \infty$ corresponds to the long-wave limit discussed in Sect.~\ref{s.long}. 

For $\beta<2/5$ there are two positive real solutions: one of them approaches zero for long wavelengths (the trivial solution of Eq.~[\ref{lw1}]) while the other corresponds to $n_0$ given in Eq.~\eqref{eee.6}. According to the classical theory of disk instabilities they are related to the secular and thermal instabilities, respectively. 
The branches corresponding to the secular and thermal modes approach each other with decreasing wavelength and merge into complex conjugate solutions (dashed lines). Two real and negative branches appear again for short wavelengths. The $1/(kH) \to 0$ limit of the upper one corresponds to Eq.~\eqref{szcx2}. The lower branch approaches $-\infty$ according to Eq.~\eqref{szsz.2}.
The same picture holds for $\beta>2/5$, except that now all real solutions
are negative, at large wavelengths one of the roots corresponds to the trivial one and approaches $0^-$ in the limit of $1/(kH) \to \infty$, while the other one approaches $n_0<0$.

For $\beta=2/5$ there are two negative real roots at short wavelengths, $F>4G_{\sigma}C_1/C_2$---in the limit $1/(kH) \to 0$ one root corresponds to  Eq.~\eqref{szcx2}, $-G_{\sigma}/C_2= -0.038$ for $\gamma=5/3$, while the other tends to $-\infty$. At $F=4G_{\sigma}C_1/C_2$ these merge into complex conjugate solutions tending to the origin of the complex plane ($n=0$) in the long-wave limit ($F\to 0$), while never becoming real for $F<4G_{\sigma}C_1/C_2$---the real part of these conjugate solutions is negative.
%++++++++++++++++++++++++++++++++++++++++++++++++++++++++++++++++
% beta=0, alpha=0.1 vs xi & tau
%+++++++++++++++++++++++++++++++++++++++++++++++++++++++++++++++++
%++++++++++++++++++++++++++++++++++++++++++++++++++++++++++++++++

\begin{figure}
\centering
 \includegraphics[width=.95\columnwidth]{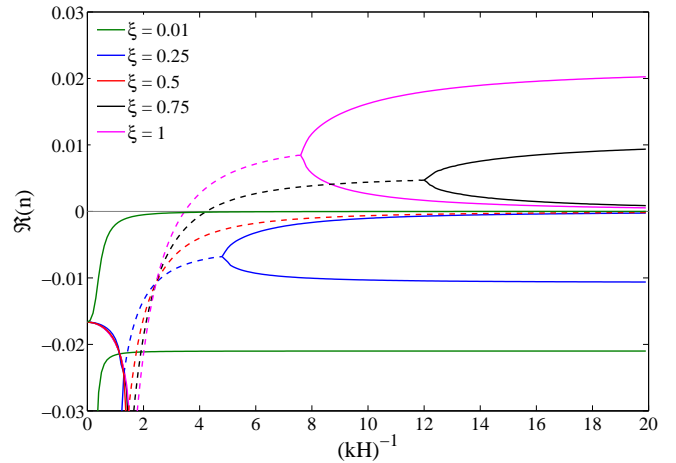}
\caption {Real part of solutions of Eq.~\eqref{disperse} as a
  function of the wave length $1/(kH)$ for $\alpha_0=0.1$, $\beta=0$, $\Omega\tau=0$ and
  various values of $\xi$. The line convention is the same as in previous figures.}
 \label{f.kHvsxi}
\end{figure}

The critical inequality \eqref{lw05} relates $\beta$ and $\xi$. Thus, for fixed $\beta$ similar effects to the ones discussed above may be obtained by varying $\xi$. In Fig.~\ref{f.kHvsxi} we plot roots of the dispersion relation, Eq.~\eqref{disperse}, for $\alpha_0=0.1$, $\beta=0$, $\Omega\tau=0$ and a few values of $\xi$. For radiation pressure dominated disks ($\beta=0$) the critical value is $\xi=1/2$. For $\xi$ higher than this value, the solutions exhibit similar behavior to those discussed above for $\beta<2/5$---two positive roots for long waves (one approaching $0^+$), a common complex conjugate branch and two negative real solutions for short waves. The latter approach $-\infty$ and the limit defined in Eq.~\eqref{szcx2}, which does not depend on $\xi$. For $\xi<1/2$ both solutions are negative for all wavelengths and approach the same limits for the shortest waves as before. The complex conjugate branch does not appear at all for the lowest presented value of $\xi=0.01$.

%++++++++++++++++++++++++++++++++++++++++++++++++++++++++++++++++
% beta=0,ksi=1 vs tau & kH
%+++++++++++++++++++++++++++++++++++++++++++++++++++++++++++++++++
%++++++++++++++++++++++++++++++++++++++++++++++++++++++++++++++++

\begin{figure}
\centering

 \subfigure{\includegraphics[width=.95\columnwidth]{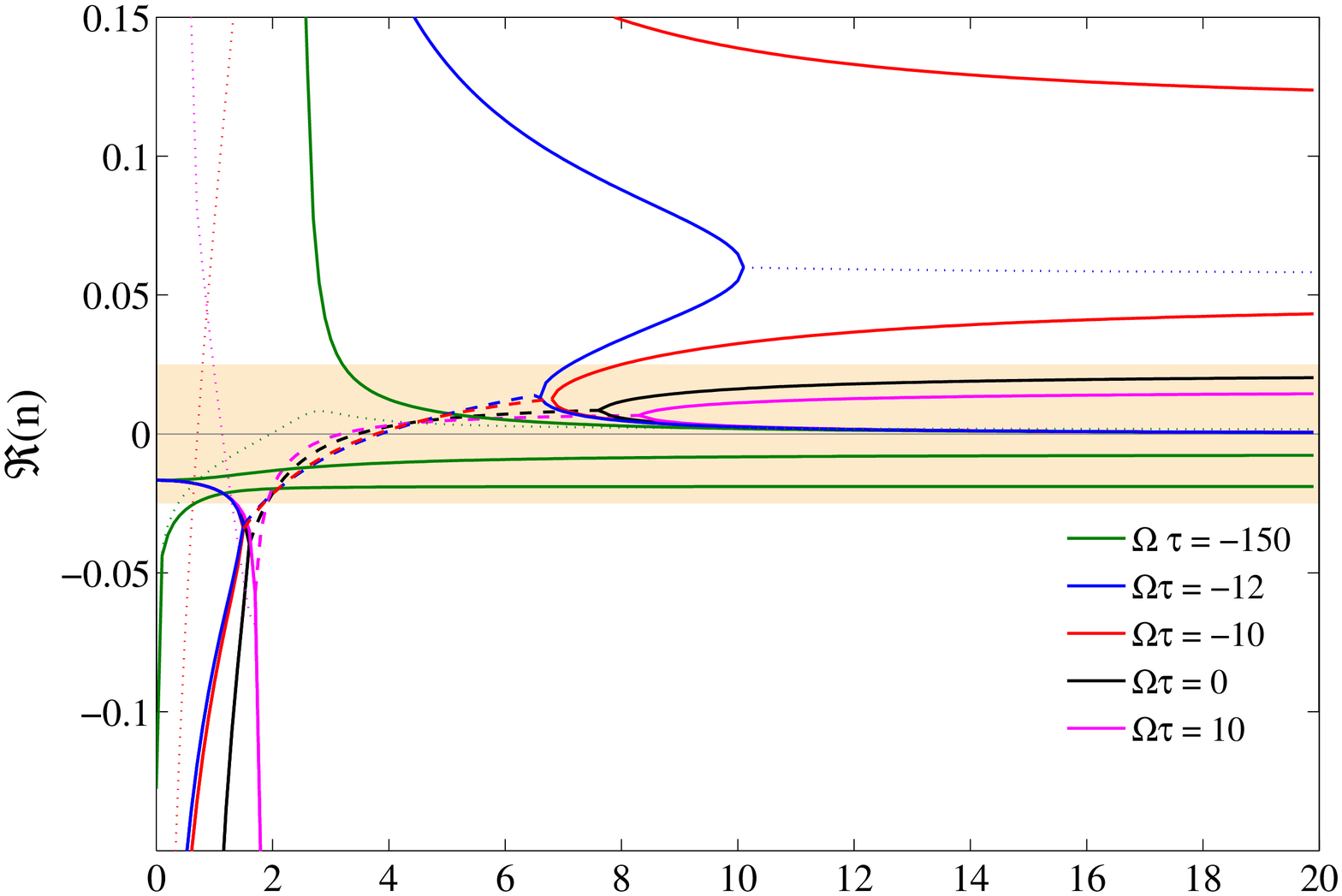}}  \\\vspace{-.75cm}
 \subfigure{\includegraphics[width=.95\columnwidth]{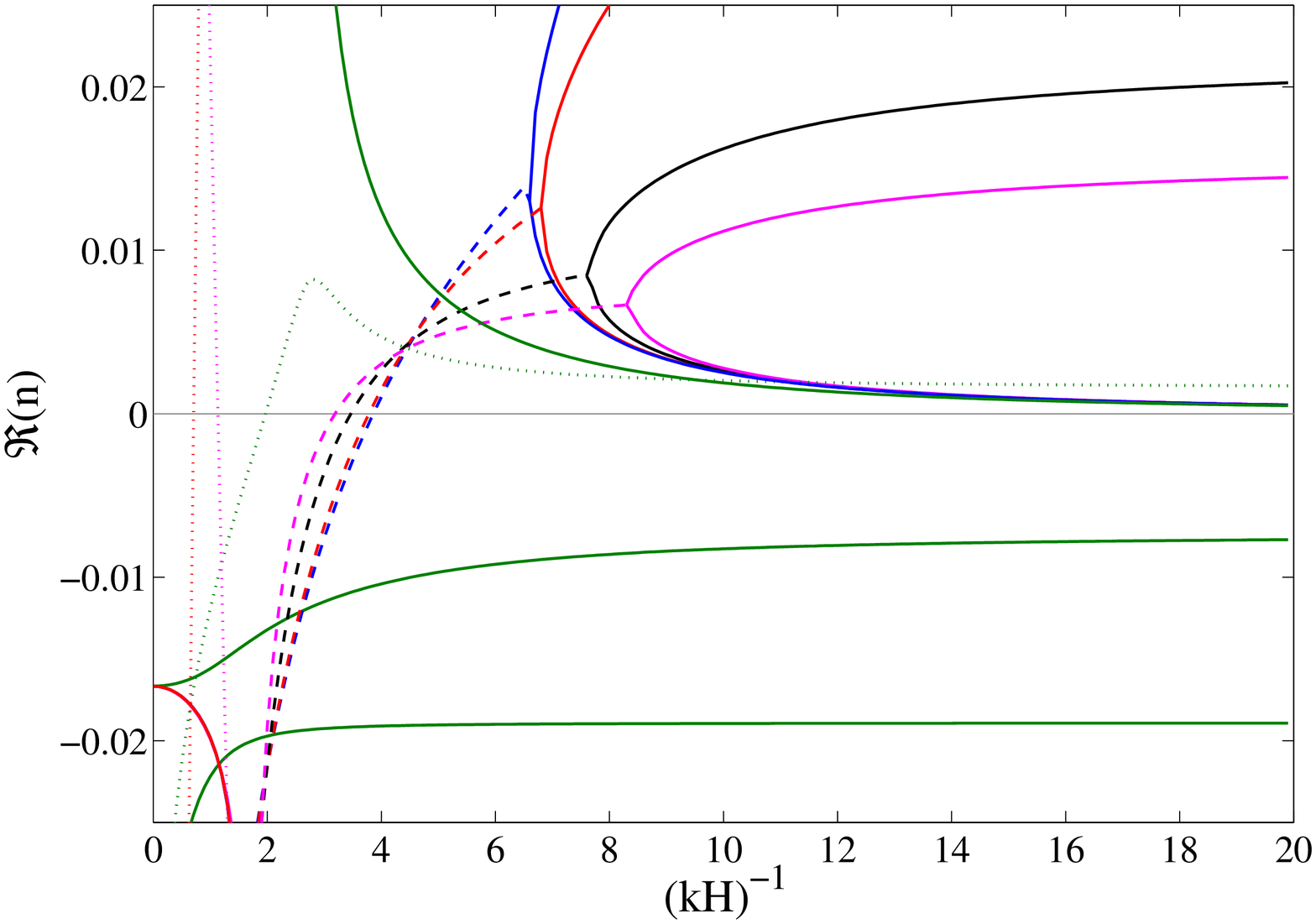}}
\caption {Real part of solutions of Eq.~\eqref{disperse} as a
  function of the wave length $1/(kH)$ for $\alpha_0=0.1$, $\beta=0$, $\xi=1$ and
  various values of the time delay $\tau$. The bottom panel zooms in
the shaded area on the top panel.
The line convention is the same as in previous figures. In this figure, we take $\gamma=5/3$.}
 \label{f.kHvstau}
\end{figure}

Fig.~\ref{f.kHvstau} presents similar plots for $\alpha_0=0.1$, $\beta=0$, $\xi=1$ and a few values of the time delay $\Omega\tau$. The long-wave limit corresponds to the solutions presented in Fig.~\ref{f.f00}. The bottom plot zooms in the shaded region in the top panel. For $\Omega\tau=0$ (black curves) we recover one of the standard cases presented in the previous plots.  Positive values of $\Omega\tau$ (e.g., magenta curves) result in two positive roots (one approaching $0^+$) for long waves. They merge into the complex conjugate branch with decreasing wavelength and again split into two negative real solutions, similarly to some of the cases discussed above.

%++++++++++++++++++++++++++++++++++++++++++++++++++++++++++++++++
% beta=0,ksi=1 vs tau & kH
%+++++++++++++++++++++++++++++++++++++++++++++++++++++++++++++++++
%++++++++++++++++++++++++++++++++++++++++++++++++++++++++++++++++

The behavior of solutions obtained with time delays ($\Omega\tau<0$) is more complicated. The number of solutions in the long-wave limit depends on the relation of the time delay to the quantities $\Omega\tau_{1}$ and $\Omega\tau_2$ (for the case presented in Fig.~\ref{f.kHvstau} these are $-11$ and $-125$, respectively, Eq.~[\ref{eee.5}]). For negative $\Omega\tau>\Omega\tau_1$ we expect in total three real roots: two positive, and one equal to zero in the long wavelength limit (the trivial one). The red curve in Fig.~\ref{f.kHvstau} ($\Omega\tau=-10$) corresponds to this case---there are three positive roots, two tending to $n\approx0.04$ and $\approx0.12$, and one approaching zero.

When $\Omega\tau<\Omega\tau_2$ (e.g., $\Omega\tau=-150$, green lines in Fig.~\ref{f.kHvstau}) there are two negative real roots of Eq.~\eqref{lw00} (compare Fig.~\ref{f.f00}) clearly visible in the bottom panel of Fig.~\ref{f.kHvstau}. The dotted green line approaching $\Re(n)\approx0.002$ in the long-wave limit corresponds to the first complex periodic solution (dotted lines in the second quadrant of Fig.~\ref{f.f00}). The third, positive, solution becomes the trivial one, $\Re(n)\to0$ for $kH\to0$.

The branches corresponding to the trivial solution of the long-wave limit leave the $\Re(n)=0$ axis and reach positive values with decreasing wavelength. For $\Omega\tau>\Omega\tau_1$ they merge with the other positive real branch corresponding to the smaller of the real roots and transform into complex conjugate branches (red dashed line). The larger real and positive root diverges with decreasing wavelength, approaching $+\infty$ according to Eq.~\eqref{sz.20}.

Once $\Omega\tau_2<\Omega\tau<\Omega\tau_1$ no real solution of Eq.~\eqref{lw00} exists. The dotted blue line (corresponding to $\Omega\tau=-12$) reflects the complex conjugate solution. At large (but finite) wavelengths, the single positive solution of Eq.~\eqref{disperse} corresponds to the trivial solution $n=0$ of the long wavelength limit, Eq.~\eqref{lw1}. However two additional real and positive solutions appear at some moderate range of wavelengths. These new branches behave similarly to the case previously discussed: one of them merges with the trivial branch, the other diverges at zero wavelength. 

For $\Omega\tau<\Omega\tau_2$ (e.g., green line) the trivial branch diverges on its own, there is only one positive root for all wavelengths and the complex conjugate branch does not appear. For the shortest wavelengths and for all negative $\Omega\tau$ there are three real solutions: two negative solutions, one with the limit given by Eq.~\eqref{szcx2} and the other approaching $-\infty$ (Eq.~[\ref{szsz.2}]), and one positive  solution approaching $+\infty$ according to Eq.~\eqref{sz.20}.

%%%%%%%%%%%%%%%%%%%%%%%%%%%%%%%%%%%%%%%%%%%%%%%%%%%%%%%%%%%%%%%%%%%%%%%%%%%%%%%
\section{Summary}
\label{s.summary}

We have derived the dispersion relation, Eq.~\eqref{disperse}, for perturbations of an accretion disk with heating that is offset in time relative to pressure perturbations. The standard $\alpha$-prescription was generalized to account for a time shift $\tau$ between the viscous stress response and perturbation of pressure, as well as for an arbitrary ratio of the corresponding perturbations $\alpha\xi$ (Sect.~\ref{s.prescription}). No restrictions were placed on the allowed gas pressure to total pressure ratio, $0\le\beta\le1$.

In the limit of long waves the number of real solutions for the perturbation growth rate, $\Omega n$, and their signs depend both on the relation between $\xi$ and $\beta$, and on the value of the time lag $-\tau$. For all cases there is one trivial solution $n=0$. For the standard case with no time lag ($\Omega\tau=0$) there is an additional real solution. It is negative if (Eq.~[\ref{lw05}]),
\be \label{lw05bis}
\xi < \frac{2(1+\beta)}{4-3\beta}.
 \ee
The same condition applies when $\Omega\tau>0$, i.e., advanced heating does not affect the appearance of the viscous instability in radiation pressure dominated disks. However, retarded heating may stabilize (or destabilize) the disk. For $\Omega\tau<0$, if inequality ~\eqref{lw05bis} is satisfied then two real roots with opposite signs exist (in addition to the trivial one). If the inequality is not satisfied two negative roots appear, but only for delays larger than a critical value, $\Omega\tau<\Omega\tau_2<0$.  The specific value of this critical $\Omega\tau_2$ is inversely proportional to $\alpha_0$ and  depends on $\beta$ and $\xi$ according to Eq.~\eqref{eee.5}. For smaller values of the time lag, the non-trivial real solutions are positive or do not exist. Properties of real solutions of the dispersion relation in the limit of long wavelengths are summarized in Table~\ref{ttt.1}.

In addition to the real roots of the dispersion equation discussed above there exist (for $\Omega\tau\neq0$) an infinite number of complex periodic solutions. Their real part, i.e., the growth rate, is opposite in sign to $\Omega\tau$ (for moderate values of $\Omega\tau$).

For very short waves there are two negative real solutions (one of the damping rates is finite, the other approaches inifnity) independently of the system parameters. For $\Omega\tau<0$ there is an additional positive root diverging to $+\infty$.

Based on these properties we may conclude that the thermal and secular branches are stable for $\Omega\tau\ge0$ if criterion \eqref{lw05bis} is satisfied. For negative $\Omega\tau$ the thermal branch is stable only if the time lag is sufficiently large. For negative $\Omega\tau$ complex solutions with a positive real part exist (Fig.~\ref{f.f00}) and therefore the issue of stability in this regime is more complicated. The growth rates of the secular branch are positive for all time lags ($\tau<0$).

In a forthcoming paper we shall  present a detailed physical discussion of the  stability of disks with retarded heating, applying our conclusions to recent results of MHD numerical simulations of radiation-pressure dominated disks.

\acknowledgements{
This work was supported in part by Polish Ministry of Science grants NN203 381436, N203 0093/1466 and N203 380336. JPL was supported
by the French Space Agency CNES. We are very grateful to
%Paola Rebusco and 
Omer Blaes for fruitful discussions.
We also thank Mateusz Janiak for helpful comments.
}

\bibliographystyle{apj}
%--------------------------------------------------------------------

\end{document}